\documentclass[manuscript,screen,nonacm]{acmart}
\AtBeginDocument{%
  }

\renewcommand\footnotetextcopyrightpermission[1]{}

\setcopyright{acmlicensed}
\copyrightyear{2018}
\acmYear{2018}
\acmDOI{XXXXXXX.XXXXXXX}
\acmConference[Conference acronym 'XX]{Make sure to enter the correct
  conference title from your rights confirmation email}{June 03--05,
  2018}{Woodstock, NY}
\acmISBN{978-1-4503-XXXX-X/2018/06}

\usepackage{multirow} 
\usepackage{listings}
\begin{document}

\title[Retrieval-Feedback-Driven Distillation and Preference Alignment for Efficient LLM-based QE]{Retrieval-Feedback-Driven Distillation and Preference Alignment for Efficient LLM-based Query Expansion}


\author{Minghan Li}
\authornote{Corresponding author.}
\affiliation{%
  \institution{School of Computer Science and Technology, Soochow University}
  \city{Suzhou}
  \state{Jiangsu}
  \postcode{215031}
  \country{China}
}
\email{mhli@suda.edu.cn}

\author{Guodong Zhou}
\affiliation{%
  \institution{School of Computer Science and Technology, Soochow University}
  \city{Suzhou}
  \state{Jiangsu}
  \postcode{215031}
  \country{China}
}
\email{gdzhou@suda.edu.cn}

\renewcommand{\shortauthors}{Li et al.}

\begin{abstract}
Large language models have recently enabled a generative paradigm for query expansion, but their high inference cost makes direct deployment difficult in practical retrieval systems. To address this issue, a retrieval-feedback-driven distillation and preference-alignment framework is proposed to transfer retrieval-friendly expansion behavior from a strong teacher model to a compact student model. Rather than relying on few-shot exemplars at inference time, the framework first leverages two complementary types of teacher-generated expansions, produced under zero-shot and few-shot prompting conditions, as supervision signals for distillation and as candidate pools for preference construction. A retrieval-metric-driven strategy is then introduced to automatically form chosen/rejected expansion pairs according to $\Delta\mathrm{nDCG@10}$, and Direct Preference Optimization is applied to explicitly align generation preferences with retrieval objectives. Experiments on TREC DL19/20/21 and MIRACL-zh show that the proposed approach preserves strong retrieval effectiveness while substantially reducing inference cost. In particular, the distilled Qwen3-4B model reaches about 97\% of the teacher (DeepSeek-685B) model's nDCG@10 performance on DL19, and remains effective on the Chinese MIRACL-zh benchmark, demonstrating strong practicality across both English and Chinese retrieval settings.
\end{abstract}




\maketitle

\section{Introduction}
Information Retrieval (IR) \cite{zhu2025large, li2024domain, li2023power} plays a fundamental role in applications such as search engines, legal case retrieval \cite{feng2024legal}, medical question answering, and intelligent customer service. Its core objective is to retrieve and rank \cite{thonet2022listwise} relevant information from massive document collections quickly and accurately according to user queries. However, traditional sparse retrieval models (e.g., BM25 \cite{robertson2009probabilistic}) mainly rely on term matching, require strong query expressiveness, and struggle to adequately capture latent semantic intent. As a result, they are often limited in scenarios involving short queries, semantic sparsity, and cross-domain retrieval. To bridge the representation gap between queries and documents, Query Expansion (QE) \cite{wang2023query2doc, li2025query} has been widely used to enhance the semantic coverage of the original query and improve retrieval ranking performance.

With the rapid development of large-scale pre-trained language models \cite{naveed2025comprehensive}, especially instruction-tuned large language models (LLMs), generative query expansion has become a research focus in recent years. Unlike traditional methods based on term reweighting or pseudo-relevance feedback \cite{datta2024deep}, LLMs can generate paragraph-level and explanatory expansion text, thereby explicitly expressing user intent, supplementing semantic background information, and providing richer matching signals for downstream retrieval. Existing studies have shown that LLM-driven generative expansion has significant advantages \cite{wang2023query2doc} and can substantially improve the effective recall capability of sparse retrieval models.

However, although large-model expansion can achieve substantial improvements in retrieval performance, its deployment cost and inference overhead make it difficult to satisfy the real-time requirements of online services. On the one hand, high-performance LLMs are large in parameter scale and slow in inference, making them difficult to deploy effectively in resource-constrained or high-concurrency retrieval scenarios. On the other hand, directly invoking extremely large models (e.g., models with more than 500B parameters) for expansion can significantly increase end-to-end retrieval latency, thereby increasing system maintenance and service costs. In contrast, small-scale large language models have advantages such as faster inference, more flexible deployment, and controllable cost. However, their generation quality, semantic reasoning ability, and domain generalization ability are all weaker than those of large models, and when directly used for query expansion, they often fail to achieve strong retrieval gains. Therefore, how to reduce inference cost while maintaining expansion quality, so that small models can acquire the expansion capability of large models, has become a key problem in current research on generative query expansion.

To address this issue, this paper proposes a preference-aligned knowledge distillation method for query expansion based on large language models \cite{matsuo2022deep, phuong2019towards}. In this method, a large model serves as the teacher model and supervised data are constructed by generating query expansions. First, supervised fine-tuning (SFT) \cite{dong2024abilities} is used to transfer the expansion capability of the large model to a small model, enabling the latter to obtain stable paragraph-level expansion ability without relying on example prompts. On this basis, we further introduce retrieval-feedback-based Direct Preference Optimization (DPO) \cite{rafailov2023direct}, which aligns preferences according to the actual contribution of different expansions to retrieval effectiveness (e.g., nDCG@10). As a result, the expansions generated by the small model are not only readable, but can also significantly improve retrieval performance. This method jointly considers expansion quality and inference efficiency, providing a practical path for deploying generative query expansion in real retrieval systems.

It should be emphasized that the goal of this paper is not to pursue the best possible effectiveness under all settings by introducing more complex multi-round generation or iterative retrieval pipelines. Instead, we study {under the strict constraints of single-pass generation and low inference cost}, how to distill the expansion capability of a strong teacher into a {directly deployable} small model, and how to use retrieval-feedback alignment to obtain stable gains on standard retrieval benchmarks.

\begin{enumerate}
    \item \textbf{A deployable distillation interface with unified prompting.}
    We design a teacher--student generation framework for query expansion, which distills the teacher model's paragraph-level expansion capability formed under different prompting conditions into a small model, enabling the student model to stably generate high-information-density expansions during inference \textbf{without relying on few-shot examples}.

    \item \textbf{Retrieval-metric-driven preference construction and alignment.}
    We propose a query-level preference sample construction strategy based on $\Delta\mathrm{nDCG@10}$, which automatically forms chosen/rejected expansion pairs, and apply DPO to \textbf{explicitly align} the student model's generation preferences with retrieval objectives, thereby improving the retrievability and ranking gains of expansions.

    \item \textbf{Complementary distillation from dual teacher expansion data.}
    We use two types of expansions generated by the teacher under zero-shot and few-shot settings (E1/E2) as complementary supervision signals, and verify their complementary roles in coverage and robustness through data-composition ablation, thereby improving the stability of distillation training.

    \item \textbf{Systematic validation and evidence of deployability.}
    We conduct extensive experiments on English TREC DL19/20/21 benchmarks and the Chinese MIRACL-zh benchmark, verifying the effectiveness of the proposed method under both sparse and dense retrieval settings. We further provide quantitative results on inference latency and GPU memory cost, supporting the core motivation of ``low-cost deployability'' and showing that the framework can be extended beyond the original English MS MARCO-style setting.
\end{enumerate}

\section{Related Work}

\subsection{Query Expansion Methods}

Query Expansion (QE) aims to alleviate the vocabulary mismatch between query expressions and document content, thereby improving both recall and ranking effectiveness in retrieval systems \cite{carpineto2012survey,li2025query, li2024evirerank, li2026efficientlongdocumentrerankingblocklevel}. Early studies mainly relied on expansion strategies based on term frequency statistics and relevance feedback, among which Pseudo-Relevance Feedback (PRF) is a representative approach. The Rocchio algorithm \cite{ye2010revisiting} constructs a new query representation by linearly combining the original query vector with feedback document vectors, while the RM3 model \cite{lavrenko2017relevance} builds a term probability distribution from feedback documents under the language modeling framework to generate richer expansion terms. However, such methods are highly dependent on the quality of the initial retrieval results. When the feedback set contains non-relevant documents, noise may be introduced, leading to query drift and making it difficult to accurately capture complex semantic intent.

With the development of deep semantic representation techniques, neural network models have gradually been introduced into query expansion tasks. The Doc2Query method enhances inverted indexes by using a sequence generation model to produce potential queries for each document, thereby improving recall capability \cite{gospodinov2023doc2query}. BERT-QE leverages deep language models to perform contextual semantic modeling over documents and select expansion content that improves relevance discrimination \cite{zheng2020bert}. These approaches partially overcome the limitations of traditional frequency-based methods, but they typically require large-scale training data, and their expansion quality is influenced by the capability of the pretrained model as well as the coverage of domain-specific corpora.

In recent years, the rapid development of large-scale instruction-tuned language models has introduced a new generative paradigm for query expansion. Unlike traditional expansion methods based on term statistics, large language models (LLMs) can generate paragraph-level, explanatory, and contextualized expansion text, enabling queries to cover richer semantic expressions and potential reasoning signals \cite{gospodinov2023doc2query, wang2023query2doc}. Prior work has shown that through prompt-based control \cite{maia2024efficient}, models can extract key information from external knowledge, linguistic co-occurrence patterns, and implicit semantic concepts, thereby improving their ability to interpret user intent. AGR \cite{chen2024analyze} adopts a three-stage prompting strategy of “analyze–generate–refine” to reduce expansion noise and focus the generated content on the core semantics of the query. MILL \cite{jia2024mill} and MUGI \cite{zhang2024exploring} further demonstrate that multi-path generation and semantic aggregation can improve the consistency and robustness of expansions, helping mitigate generation drift and semantic dilution. CSQE \cite{lei2024corpus} proposes a corpus-guided expansion approach that introduces document evidence during the generation process, reducing potential factual hallucinations produced by LLMs and improving alignment between generated expansions and the target corpus domain. SU-RankFusion \cite{li2026dual} further addresses the brittleness of prompt-only generative QE by decoupling stable behavioral control from diversified user-side prompts and aggregating the resulting ranked lists with lightweight fusion, thereby improving the robustness of zero-grounding query expansion.

Overall, LLM-driven generative query expansion has demonstrated strong effectiveness in scenarios such as cross-domain retrieval and short-query understanding. By generating explanatory expansion text, these methods enrich query semantics and improve both recall and ranking performance. Nevertheless, large models still face significant challenges in inference latency \cite{husom2025sustainable}, computational cost, and real-world deployment. Directly applying such models for online expansion can substantially increase system overhead, making it difficult to satisfy the requirements of high-concurrency and low-latency applications. In contrast, small-scale language models offer advantages such as faster inference, flexible deployment, and lower computational cost. However, their generation and semantic modeling capabilities are limited, and directly using them for query expansion often fails to fully capture latent semantic relationships within queries, leaving a substantial gap in expansion quality compared to large models. Therefore, improving the expansion capability of small models while maintaining low inference cost remains a key challenge. A natural direction, which is explored in this work, is to use large models as the source of capability and transfer their expansion patterns to smaller models through knowledge distillation, combined with preference alignment based on generation behavior and retrieval metrics, enabling small models to approach the expansion quality of large models while preserving high inference efficiency.

\subsection{Knowledge Distillation and Reinforcement Learning Alignment}

Although large language models exhibit strong expressive capability for query expansion, directly using them introduces challenges such as high computational cost, long inference latency, and expensive deployment requirements. Consequently, transferring the capability of large models to smaller models has become an important research direction. Knowledge Distillation (KD) \cite{phuong2019towards} enables a student model to learn from a teacher model by transferring output distributions, hidden representations, or behavioral patterns, allowing the student model to achieve performance close to the teacher while maintaining lower complexity. In natural language generation tasks, Supervised Fine-Tuning (SFT) is commonly used to train the student model to reproduce the textual style and information structure generated by the teacher, thereby obtaining stable generation capability. However, SFT primarily aligns generation distributions and cannot guarantee consistent improvements in downstream retrieval metrics such as nDCG@10, MAP, or Recall. This leads to the so-called mismatch between “readability” and “retrievability”.

To reduce the gap between generation quality and task objectives, preference modeling and reinforcement learning (RL) methods have been introduced to align the generation behavior of large language models. Early approaches based on Reinforcement Learning from Human Feedback (RLHF) \cite{ouyang2022training} rely on explicit reward models to guide generation behavior, but they are computationally expensive and require human annotations. More recently, Direct Preference Optimization (DPO) \cite{rafailov2023direct} has been proposed as a more efficient alternative that directly optimizes generation preferences using pairwise preference samples without the need to train a reward model. Such approaches provide a practical solution to the problem of strong generation capability but insufficient task alignment, by constraining the model output space toward directions that better satisfy downstream task objectives.

Beyond general reinforcement learning frameworks, some studies have attempted to explicitly incorporate downstream task feedback into the generation process so that expansions can directly serve retrieval or question answering objectives. For instance, keyword generation in search advertising leverages implicit behavioral signals such as click logs to learn query–keyword matching relationships and generate expansion terms that are more beneficial for retrieval matching, particularly for long-tail queries \cite{lee2018rare}. In retrieval-based question answering, question rewriting can be formulated as a sequence generation task, where deep reinforcement learning jointly models language quality and answer retrievability or relevance as reward signals to improve answer retrieval performance \cite{liu2019generative}. These studies embody the idea of constraining generation outputs using task-level feedback. However, their task settings and feedback sources differ from those of general Web retrieval query expansion: the former typically rely on user behavior data such as click logs, while the latter focus on reinforcement learning strategies for question rewriting in QA scenarios.

Recent work has also explored combining distillation with preference or feedback signals to improve the efficiency and alignment of query expansion. RADCoT \cite{lee2024radcot} generates explanatory or reasoning-style expansions using a teacher model and distills them into a retrieval-enhanced student model. During inference, the student model conditions on retrieved evidence passages to generate expansion text that is then concatenated with the query. ExpandR \cite{yao2025expandr} adopts a system-level approach that alternately optimizes the retriever and the generator, encouraging the generator to produce expansions that are more useful for the retriever while improving the retriever’s ability to exploit expansion signals. In general, these methods fall into the paradigms of retrieval-augmented generation or system-level joint optimization. They typically require additional retrieval context or joint training with a retriever, resulting in training and inference budgets closer to end-to-end system construction.

In contrast, this work focuses on learning a query expansion model that supports single-pass generation and low-latency deployment. Under the constraints of a fixed retriever and a unified zero-shot deployment interface, we transfer the paragraph-level expansion capability of a high-cost teacher model to a small model through knowledge distillation. We further construct pairwise preference supervision based on retrieval evaluation metrics to align the generation behavior of the student model. This two-stage training mechanism enables systematic evaluation of improvements in both retrieval effectiveness and efficiency without introducing multi-round retrieval–generation loops or joint retriever training.

In summary, although existing research has evolved from statistical expansion methods toward LLM-based semantic generation, a gap still exists between generated content and retrieval metrics. Knowledge distillation provides an effective approach for reducing inference cost and enabling local deployment, while preference optimization offers a direct mechanism for aligning generation behavior with downstream retrieval objectives. Therefore, developing a single-pass, low-latency, deployable expansion model through distillation and alignment remains an important research direction.

\section{Preference-Aligned Knowledge Distillation for Query Expansion Based on Large Language Models}

\begin{figure*}[t]
\centering
\includegraphics[width=0.95\textwidth]{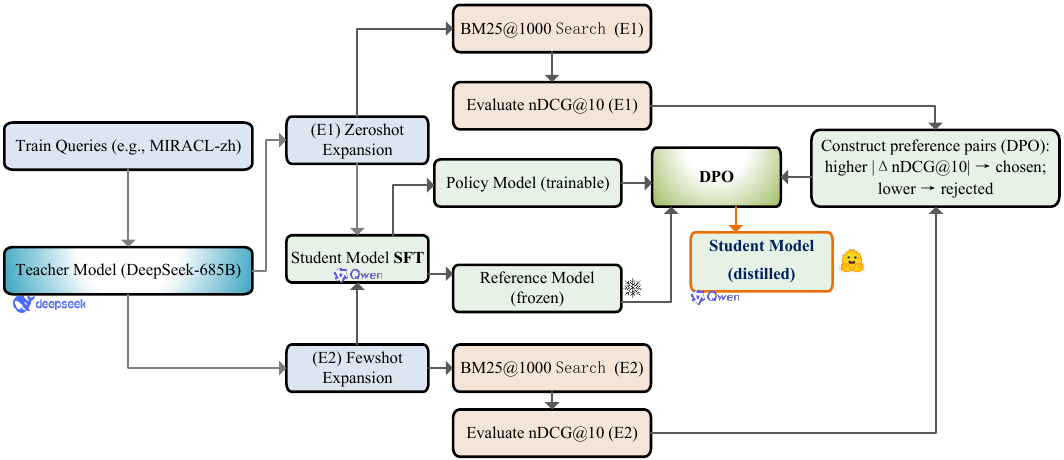}
\caption{Overall system architecture of the proposed method.}
\label{fig:framework}
\end{figure*}

The overall architecture of the proposed method is shown in Fig.~\ref{fig:framework}. In the following, we describe the method from the perspectives of supervised data construction, supervised fine-tuning, and preference alignment.

\subsection{Supervised Data Construction with Large Language Models}

As discussed above, in order to enable smaller large language models to perform query expansion more effectively, this work constructs teacher data using a larger large language model. Specifically, for queries in the information retrieval domain, we use training queries and employ a large language model to generate query expansions. The default large language model used in this work is DeepSeek-V3.1-Base\footnote{\url{https://huggingface.co/deepseek-ai/DeepSeek-V3.1-Base}} (685B). Concretely, we first call the API provided by DeepSeek to generate expansions for the queries. 
In Section~\ref{sec:otherOlmo32Teacher}, we additionally report results with another teacher model.

We choose DeepSeek-V3.1-Base (685B) as the teacher model for three reasons. First, it exhibits strong instruction-following ability and long-form generation capability, allowing it to produce stable paragraph-level expansions with high information density under a unified prompting setup, which is important for constructing high-quality distillation data. Second, its relatively stable serving interface makes it feasible to generate expansions for a large number of training queries with modest engineering overhead, thereby covering a broader query distribution and improving the generalization ability of the distilled student. Third, the proposed two-stage framework of distillation and retrieval-feedback-driven alignment is not specific to DeepSeek. To demonstrate this, we additionally introduce another open-source teacher model in supplementary experiments and observe consistent effectiveness trends.

We adopt a ``dual'' generation strategy: for each training query, two expansions are generated simultaneously, namely the {zero-shot} expansion (E1) and the {few-shot} expansion (E2). Our generation method follows the strategy below: for the same query, we first generate E1 without examples, and then generate E2 by appending a small number of examples to the same prompt template. In this way, ``one query, two answers'' automatically forms two types of data, which can later be used for both imitation-based supervised distillation and preference-based alignment distillation. To ensure data quality and consistency, the generated expansion text is normalized before being written out, including the removal of newline characters, tab characters, and other special symbols that may cause abnormal text structure.

\subsubsection{Prompt Design}

Based on instruction-tuned large language models, we adopt a two-level prompt structure and divide the prompt into {system} and {user} parts. The {system} part is used to define the overall behavioral constraints and output style of the model, while the {user} part serves as the task-specific instruction, taking {Query: \{Q\}} as input, where $Q$ is the original query text. The model is required to generate an explanatory paragraph of 60--100 words, whose content should directly cover the user's intent while maintaining semantic coherence and information density.

\subsubsection{Zero-shot Prompt Template}

\begin{figure*}[t]
\centering
\begin{minipage}{0.95\textwidth}
\lstset{
  basicstyle=\ttfamily\footnotesize,
  breaklines=true,
  columns=fullflexible,
  frame=single,
  keepspaces=true
}
\begin{lstlisting}
system:
  You are an assistant that generates detailed passages to answer search queries. Your responses should be informative, directly address the query, and provide comprehensive explanations or solutions.

user:
  Query: {Q}
  Please write a passage (60-100 words) that answers it.
\end{lstlisting}
\end{minipage}
\caption{Zero-shot prompt template.}
\label{fig:prompt_zero}
\end{figure*}

In the zero-shot setting, no examples are provided, and generation is directly driven by the query. For English, the template is shown above, where {\{Q\}} denotes the original query to be expanded.
For Chinese queries, we use a Chinese counterpart of the same template, requiring the model to generate a fluent Chinese expansion passage with the same length constraint and task objective. To be specific, the last sentence is changed to ``Please write a passage (60-100 words) in Chinese that answers it''.

\subsubsection{Few-shot Prompt Template}

In the few-shot setting, multiple ``query-expansion examples'' are added to the {user} part to demonstrate paragraph structure, writing style, and semantic elaboration patterns; the target query is then given with the same 60--100 word constraint. The example set used in this work consists of four classic examples from the Query2Doc paper \cite{wang2023query2doc}. These four examples do not overlap with the training or test queries of any experimental dataset used in this paper, and are used only to demonstrate the prompt format. All of them follow a strongly constrained ``query-explanatory paragraph'' pattern, which can effectively guide the model to generate more retrievable expansion text.

\begin{figure*}[t]
\centering
\begin{minipage}{0.95\textwidth}
\lstset{
  basicstyle=\ttfamily\footnotesize,
  breaklines=true,
  columns=fullflexible,
  frame=single,
  keepspaces=true
}
\begin{lstlisting}
system:
  You are an assistant that generates detailed passages to answer search queries. Your responses should be informative, directly address the query, and provide comprehensive explanations or solutions.

user:
  Query: what state is this zip code 85282
  Passage: Welcome to TEMPE, AZ 85282. 85282 is a rural zip code in Tempe, Arizona. The population is primarily white and mostly single. At $200,200 the average home value here is a bit higher than average for the Phoenix-Mesa-Scottsdale metro area, so this probably is not the place to look for housing bargains. 85282 Zip code is located in the Mountain time zone at 33 degrees latitude (Fun Fact: this is the same latitude as Damascus, Syria) and -112 degrees longitude.
  Query: why is gibbs model of reflection good
  Passage: In this reflection, I am going to use Gibbs (1988) Reflective Cycle. This model is a recognised framework for my reflection. Gibbs (1988) consists of six stages to complete one cycle which is able to improve my nursing practice continuously and learning from the experience for better practice in the future. In conclusion of my reflective assignment, I mention the model that I chose, Gibbs (1988) Reflective Cycle as my framework of my reflective. I state the reasons why I am choosing the model as well as some discussion on the importance of doing reflection in nursing practice.
  Query: what does a thousand pardons means
  Passage: Oh, that is all right, that is all right, give us a rest; never mind about the direction, hang the direction - I beg pardon, I beg a thousand pardons, I am not well today; pay no attention when I soliloquize, it is an old habit, an old, bad habit, and hard to get rid of when one's digestion is all disordered with eating food that was raised forever and ever before he was born; good land! A man cannot keep his functions regular on spring chickens thirteen hundred years old.
  Query: what is a macro warning
  Passage: Macro virus warning appears when no macros exist in the file in Word. When you open a Microsoft Word 2002 document or template, you may receive the following macro virus warning, even though the document or template does not contain macros: C:\<path>\<file name> contains macros. Macros may contain viruses.
  Query: {Q}
  Please write a passage (60-100 words) that answers it.
\end{lstlisting}
\end{minipage}
\caption{Few-shot prompt template with four in-context examples.}
\label{fig:prompt_few}
\end{figure*}

For consistency across datasets, we keep the same four few-shot exemplars originally used in the English Query2doc-style prompting setup, even when the target query is in Chinese. That is, MIRACL-zh queries are prompted with the same English few-shot demonstrations.
For Chinese queries, the last sentence is simply changed to ``Please write a passage (60-100 words) in Chinese that answers it''.

\subsection{Supervised Fine-Tuning of the Student Model with Fewer Parameters}\label{sec:sft}

After obtaining the query expansion corpus generated by the teacher model (DeepSeek), we perform supervised fine-tuning (Supervised Fine-Tuning, SFT) on a smaller student model, i.e., a large language model with fewer parameters, so as to reproduce the teacher's language organization and information elaboration capability for the ``query $\!\to\!$ paragraph-level expansion'' task without relying on external APIs. To reduce training and deployment cost, we adopt parameter-efficient fine-tuning (Parameter-Efficient Fine-Tuning, PEFT) strategies, namely LoRA \cite{hu2022lora} / QLoRA \cite{dettmers2023qlora}, and update only a small number of low-rank adaptation matrices, so that distillation-based learning can also be completed in a single-GPU environment.

\subsubsection{Prompt Consistency and Input-Output Format}

During the SFT stage, we strictly reuse the conversational prompt structure used in teacher data construction. The input consists of {system} and {user} parts, where the {user} part contains the real query and the length constraint (``60--100 words, single paragraph, direct answer''). Different from the teacher stage, SFT training {no longer concatenates examples}, i.e., the student input adopts a {zero-shot} prompt, while the teacher output (target text) comes from the expansion paragraph generated by DeepSeek. Each query corresponds to two expansions (Zero/Few), which together constitute the supervision targets. This enables the student, under a zero-shot prompt, to still inherit the structured writing style and higher information density induced by the few-shot context {through the target text}.

\subsubsection{Training Objective and Loss Function}

Let $x$ denote the input sequence containing the system and user prompts, and let $y = (y_1, \dots, y_T)$ denote the target expansion paragraph generated by the teacher model. In the SFT stage, we adopt the standard autoregressive language modeling objective:
\begin{equation}
\mathcal{L}_{\mathrm{SFT}}(\theta)
=
- \sum_{t=1}^{T} \log p_{\theta}(y_t \mid x, y_{<t}),
\label{eq:sft_loss}
\end{equation}
We model the complete paragraph in an end-to-end manner, i.e., we minimize the token-level negative log-likelihood so that the conditional distribution of the student model is aligned with that of the teacher model.

SFT enables the student to obtain stable paragraph-level expansion capability and an initial alignment with the teacher's style. However, its optimization objective is still mainly oriented toward the language distribution itself, and has not yet explicitly transformed ``readable expansions'' into ``retrieval-effective expansions''. Therefore, the distillation method needs to be further optimized. In the next subsection, we introduce reward construction based on preference feedback and retrieval signals, and further perform retrieval-aware policy optimization on top of the SFT initialization, so that generation quality and downstream metrics such as nDCG@10 form a consistent objective coupling.

\subsection{Preference-Aligned Knowledge Distillation}\label{sec:dpo}

After obtaining stable paragraph-level expansion capability through SFT, we further apply Direct Preference Optimization (DPO) to perform retrieval-aware preference alignment on the student model, so as to explicitly align the generation distribution toward ``expansions that better improve retrieval metrics''. Intuitively, we feed two expansions generated for the same query under different generation strategies, such as zero-shot and few-shot, into the same retrieval and evaluation pipeline. If one of them yields a higher nDCG@10, it is regarded as the ``preferred answer'' ({chosen}), while the other is treated as the ``non-preferred answer'' ({rejected}). Based on this, pairwise preference samples are constructed and DPO training is performed.

\subsubsection{Preference Data Construction (Generation of DPO Training Pairs)}\label{sec:dpo-data}

\textbf{Source of candidate expansions and per-query evaluation.}
For each query, we use the two teacher-generated expansions obtained under zero-shot and few-shot prompting, denoted as $E_1$ and $E_2$, as candidate expansions. Each candidate is concatenated with the original query and evaluated using the same retriever as in the final test setting (e.g., BM25), under a unified index and identical retrieval parameters. We then compute the per-query retrieval effectiveness of the two expanded queries in terms of nDCG@10. In implementation, this process is carried out in a batched manner, following the pipeline of batch retrieval $\rightarrow$ per-query split $\rightarrow$ per-query evaluation, which avoids the inefficient alternative of loading the index separately for each query.

\smallskip
\textbf{Margin threshold and preference pair construction.}
To improve the reliability of preference labels, we introduce a minimum performance gap threshold $\Delta$. Only when the two candidate expansions for the same query satisfy
\begin{equation}
\big|\mathrm{nDCG@10}(E_1)-\mathrm{nDCG@10}(E_2)\big| \ge \Delta
\label{eq:delta}
\end{equation}
are they retained for DPO training; otherwise, the pair is regarded as having no clear preference and is discarded. For each retained query, the expansion with the higher nDCG@10 score is assigned as the \emph{chosen} response, while the other is assigned as the \emph{rejected} response. The resulting chosen/rejected pair is then combined with the original query to form one DPO training instance.

\subsubsection{DPO Objective and Training Setup}\label{sec:dpo-objective}

\textbf{Objective function.}
Let $\pi_{\theta}$ denote the policy (student) distribution to be optimized, and let $\pi_{\mathrm{ref}}$ denote the frozen reference distribution, which is loaded from the same backbone model and kept fixed. Let $x$ denote the zero-shot conditional input sequence consistent with the SFT stage, constructed by concatenating the system and user prompts, and let $y^{+}$ and $y^{-}$ denote the chosen and rejected expansions, respectively. DPO maximizes the following empirical objective:
\begin{equation}
\begin{aligned}
\mathcal{J}_{\mathrm{DPO}}(\theta)
= \mathbb{E}_{(x, y^+, y^-)} \Big[
\log \sigma\!\Big(
\beta \Big(
\ell_{\theta}(y^+ \mid x)-\ell_{\theta}(y^- \mid x) \\
\qquad {}-\ell_{\mathrm{ref}}(y^+ \mid x)+\ell_{\mathrm{ref}}(y^- \mid x)
\Big)\Big)
\Big].
\end{aligned}
\label{eq:dpo}
\end{equation}
where $\ell_{\theta}(y \mid x)=\log\pi_{\theta}(y \mid x)$ and $\ell_{\mathrm{ref}}(y \mid x)=\log\pi_{\mathrm{ref}}(y \mid x)$ denote the conditional log-likelihood of the policy model and the reference model, respectively; $\sigma(\cdot)$ denotes the sigmoid function, and $\beta > 0$ is the temperature coefficient. This objective encourages the student model, relative to the reference distribution, to assign higher likelihood to better expansions and suppress preference for inferior expansions under the same input condition, thereby achieving preference alignment in the sense of retrieval feedback.

\textbf{Policy model and reference model.}
In DPO, we maintain a trainable policy model $\pi_{\theta}$ and a frozen reference model $\pi_{\mathrm{ref}}$. Here, $\pi_{\theta}$ is initialized from the student model obtained after the SFT stage and is further optimized during DPO, while $\pi_{\mathrm{ref}}$ is taken as a frozen copy of the same backbone model, and only $\pi_{\theta}$ is updated during training while $\pi_{\mathrm{ref}}$ remains unchanged. The frozen reference model provides a stable distributional baseline, which helps suppress degeneration caused by excessive deviation from the backbone distribution during preference amplification, such as mode collapse and redundant generation. Meanwhile, the contrast term $\log \pi_{\theta}(y \mid x)-\log \pi_{\mathrm{ref}}(y \mid x)$ can be viewed as a soft constraint, so that preference improvement is reflected as a relative and controllable adjustment rather than an unconstrained absolute shift. During training, the two models share the same tokenizer and prompt format, and compute conditional likelihoods on the same batch of samples to construct the contrastive objective in Eq.~\eqref{eq:dpo}.

Compared with pure SFT, DPO further pushes expansions with ``good readability'' toward the direction of being ``effective for retrieval''. Its preference signal is directly derived from the per-query change in downstream retrieval metrics, and the $\Delta$-margin filtering ensures the robustness of the training pairs. Since the prompt remains consistent during both SFT and DPO, i.e., the zero-shot system+user format, the model can reproduce a generation distribution with ``high nDCG@10 preference'' during deployment without relying on external examples, thereby achieving stable gains in actual retrieval.

\section{Experiments}

\subsection{Datasets}

For the test sets, we use the TREC 2019 Deep Learning passage ranking dataset \cite{craswell2020overview}, the TREC 2020 Deep Learning passage ranking dataset\footnote{\url{https://microsoft.github.io/msmarco/TREC-Deep-Learning-2020.html}}, the TREC 2021 Deep Learning passage ranking dataset\footnote{\url{https://microsoft.github.io/msmarco/TREC-Deep-Learning-2021.html}}, and the MIRACL-zh benchmark\footnote{\url{https://huggingface.co/datasets/miracl/miracl}}  (Chinese subset) \cite{zhang2023miracl}. The retrieval source for TREC 2019 and 2020 is MS MARCO v1, whereas the retrieval source for TREC 2021 is MS MARCO v2, which contains a larger document collection. We additionally include MIRACL-zh to examine whether the proposed framework remains effective in a Chinese retrieval setting beyond the original English MS MARCO-style benchmarks. The statistics of all datasets are summarized in Table~\ref{tab:data_stats}, covering both English and Chinese retrieval scenarios across different corpus scales.

\begin{table}[htbp]
\caption{Statistics of datasets used in the experiments.}
\label{tab:data_stats}
\centering
\begin{tabular}{lccc}
\toprule
Dataset & \# Test Queries & Corpus Type & Corpus Size \\
\midrule
TREC DL19 & 43  & Web (MS~MARCO v1) & about 8.8M \\
TREC DL20 & 54  & Web (MS~MARCO v1) & about 8.8M \\
TREC DL21 & 53  & Web (MS~MARCO v2) & about 138M \\
MIRACL-zh  & 393 (dev as test) & Wikipedia (Chinese) & 4,934,368 \\
\bottomrule
\end{tabular}
\end{table}

Teacher query expansion data are constructed in a benchmark-specific manner while keeping the same generation and alignment pipeline. For the English setting, we use the MS MARCO training queries associated with the TREC DL benchmarks, from which 8,000 queries are sampled to generate teacher expansions. For the Chinese setting, we use the official MIRACL-zh training queries, which contain 1,312 queries. In both settings, DeepSeek (685B) is employed to generate candidate expansions, which are then used to build the data for supervised fine-tuning and the subsequent retrieval-metric-driven preference alignment.

\subsection{Experimental Setup}\label{sec:exp}

\subsubsection{Expansion Generation Setup}

During training data construction, we generate zero-shot expansions and few-shot expansions for each query, respectively. The prompt template follows a unified {system+user} dialogue format. Specifically, zero-shot contains only the user query, while few-shot additionally includes four examples. When using the DeepSeek chat-style interface to generate teacher expansions, we do not explicitly set temperature, nucleus sampling, or maximum generation length in the API call. Therefore, the default server-side sampling hyperparameters are used (the temperature and {top-p} take the provider defaults; the maximum generation length is also constrained by the interface default and the model limit, although in subsequent training and testing the teacher model inputs are truncated to $\leq 128$ tokens). To ensure comparability across different models, all student models, including those before distillation, after SFT, and after DPO, use the same zero-shot dialogue template and input format as the teacher model during the expansion generation stage at inference time, together with deterministic decoding, so that performance differences can be attributed to the distillation and preference alignment processes themselves.

\subsubsection{Model and Training Configuration}

The teacher model is the large model DeepSeek-V3.1 (685B) \cite{deng2025exploring}, which is used to generate query expansion text, including both zero-shot and few-shot branches. The student model is {Qwen3-4B-Instruct}\footnote{\url{https://huggingface.co/Qwen/Qwen3-4B-Instruct-2507}}. During both training and inference, the student model is loaded with QLoRA (Quantized Low-Rank Adaptation): the backbone parameters are quantized to 4-bit NF4 and fully frozen, and only the LoRA low-rank adaptation matrices are trainable. During training and inference, the query expansion length is limited to 128 tokens. The tokenizer adopts Qwen3's Byte-Level BPE tokenization scheme \cite{wang2020neural}, with left padding, and the model's {eos\_token} is used as both the padding token and the termination token, so as to keep the prompt format consistent between training and inference.

\subsubsection{Supervised Fine-Tuning (SFT)}

SFT uses $\langle\!\text{query}, \text{query expansion}\!\rangle$ as supervised samples and learns the teacher distribution under a unified {zero-shot} prompt, enabling the student to {implicitly} absorb the paragraph organization and lexical coverage induced by few-shot prompting without relying on examples at inference time. Considering deployment-side GPU memory and latency constraints, the student backbone is loaded in 4-bit quantized form (following the QLoRA paradigm), and low-rank adapters are injected into key paths such as attention projections and feed-forward bottlenecks. During training, only the newly introduced LoRA parameters are updated, allowing stable convergence on a single GPU. The optimization objective is standard cross-entropy, and only the LoRA parameters are updated. The main hyperparameters are as follows: batch size is 2, gradient accumulation steps are 4 (effective batch size 8), the learning rate is set to $2\times10^{-5}$, the AdamW optimizer and a Cosine learning-rate schedule are used, and warmup is applied during the first 10\% of training steps; the number of training epochs is 2; the LoRA rank is $r{=}4$, and the LoRA scaling factor is $\alpha{=}32$; weight quantization uses 4-bit NF4, and the computation precision is FP16. The LoRA adapter obtained from SFT is then used to initialize the policy model for DPO.

\subsubsection{Preference Alignment (DPO)}

To make the generated expansions better match the retrieval objective, we adopt Direct Preference Optimization (DPO) for preference alignment. For the Zero/Few expansions of the same query, we perform BM25@1000 retrieval separately (i.e., the top 1000 documents are taken after retrieval for each query) and evaluate them with $\mathrm{nDCG@10}$. If they satisfy the threshold $\lvert \Delta \mathrm{nDCG@10} \rvert \ge 0.01$, the better-performing one is labeled as {chosen} and the lower-performing one as {rejected}, forming a preference training pair. Training adopts a parameter-efficient strategy based on 4-bit quantization + LoRA. In DPO training, two LoRA configurations are loaded simultaneously: one is the {trainable policy model} $\pi_{\theta}$, initialized from the SFT adapter; the other is the {frozen reference model} $\pi_{\mathrm{ref}}$, which shares the same backbone as the policy model but {does not load} the SFT adapter, and is fully frozen under 4-bit precision and used only for the contrastive term. The temperature coefficient is $\beta{=}0.05$. During inference, only the {zero-shot} prompt is used to generate expansions, which are then concatenated with the original query and retrieved using BM25. All training and inference are conducted on a single Nvidia V100-32GB GPU. The learning rate is set to $2\times 10^{-6}$, the AdamW optimizer and a Cosine learning-rate schedule are used, the warmup ratio is 0.05, the batch size is 2, 4 gradient accumulation steps are used (effective batch size 8), and the number of training epochs is 2.

\subsubsection{Baselines}

We construct the baselines and our method around two aspects: whether query expansion is applied and the source/training scheme of expansion generation. The compared methods are as follows.

\begin{description}
    \item[Retrieval without expansion (BM25):]
    Only the original query is used for BM25 retrieval, serving as the most basic term-matching baseline.

    \item[BM25+Rocchio:]
    On top of BM25, Rocchio pseudo-relevance feedback \cite{ye2010revisiting} is applied to linearly update the query vector, in order to evaluate the effect of traditional PRF on recall and ranking (the implementation parameters are consistent with the defaults of Anserini \cite{yang2018anserini}).

    \item[Original model expansion (zero-shot):]
    The student backbone model (Qwen3-4B-Instruct) generates paragraph-level expansion text under a unified zero-shot prompt, and the expansion is concatenated with the original query for BM25 retrieval. This is used to measure the direct expansion capability of the model \emph{without distillation/alignment training}.

    \item[Query2Doc (few-shot):]
    Following the few-shot prompting style of Query2Doc~\cite{wang2023query2doc}, a small number of examples are added into the prompt to guide the model to generate more retrieval-friendly expansion text, which is then concatenated with the original query for BM25 retrieval. This method is used to compare against an expansion strategy that is training-free but relies on long prompts.

    \item[SFT student expansion (ours):]
    Using expansions generated by the teacher model as supervision signals, we perform QLoRA-based supervised fine-tuning on the student model (Qwen3-4B-Instruct), so that the student can learn the teacher's paragraph organization and lexical coverage without relying on example inputs.

    \item[DPO-only (ours):]
    Without SFT-based supervised distillation, the model is initialized directly from the student backbone and trained with DPO alignment using preference pairs constructed from retrieval metrics. This setting is used to examine the effectiveness and limitations of preference alignment only in the absence of supervised signals.

    \item[SFT+DPO student expansion (ours):]
    Built on top of SFT initialization, chosen/rejected preference pairs are constructed according to the relative contribution of expansions to retrieval performance (e.g., $\mathrm{nDCG@10}$), with threshold $\Delta = 0.01$, and DPO is used for preference alignment training so as to further match the retrieval objective.
\end{description}

Regarding the comparison scope with recent multi-round LLM-QE methods:  
Some recent LLM-QE methods, such as AGR \cite{chen2024analyze}, MILL \cite{jia2024mill}, MUGI \cite{zhang2024exploring}, and CSQE \cite{lei2024corpus}, usually include components such as multi-round generation, iterative retrieval, reflective rewriting, and multi-branch fusion. Their research objective is often system-level effectiveness improvement, rather than optimizing a directly deployable expansion model under the constraints of single-pass generation and low-latency deployment. As a result, it is difficult to fairly isolate the contribution of "the expansion model itself", and such settings would also deviate from the lightweight distillation and preference alignment problem studied in this paper. Therefore, under the {single-pass generation} setting, this paper mainly compares three categories of baselines:  
(1) training-free expansion methods (e.g., Query2Doc and traditional feedback models);  
(2) student models at different training stages (zero-shot/SFT/DPO-only/SFT+DPO);  
(3) the teacher model as an upper bound,  
in order to verify the effectiveness of the proposed two-stage aligned distillation framework and the source of its gains.

Through the above comparative experiments, we can systematically analyze the role and contribution of different strategies and training stages. Specifically, the original model expansion and Query2Doc are used to evaluate the generative expansion capability without additional parameter updates, and to compare the impact of zero-shot and few-shot prompting on expansion quality and deployment cost. SFT transfers the teacher model's expansion patterns through distillation, improving the stability and retrievability of expansions while maintaining inference efficiency. Furthermore, DPO-only and SFT+DPO are used to characterize the gain difference of "preference alignment" with and without supervised distillation initialization, and to examine whether the retrieval-feedback-based alignment mechanism can continuously improve retrieval performance.

\subsubsection{Retrieval and Evaluation Pipeline}

The generated expansion is concatenated with the original query and then used for sparse retrieval with BM25, where the number of candidate documents is set to 1000. Retrieval is implemented with the Anserini information retrieval toolkit \cite{yang2018anserini}. Using pre-indexed document collections, queries or expanded queries can be retrieved through the Java toolkit, and all parameters are set to their default values.

For the evaluation of DeepSeek expansion pairs and the experimental baselines in this paper, the combination of the original query and expanded query follows Query2Doc \cite{wang2023query2doc}:
\[
q_{e} = \underbrace{q \; || \; q \; || \; \ldots}_{\text{repeated 5 times}} \; || \; q^+
\]
where $q$ denotes the original query, $q^{+}$ denotes the paragraph-level expansion generated by the teacher/student model, and "$||$" denotes string concatenation in sequential order. Repeating $q$ five times is intended to explicitly amplify the weight of the original query terms under the BM25 framework, so as to avoid the expansion text overshadowing the original retrieval intent in terms of term frequency.

The evaluation metrics are standard information retrieval metrics: $\mathrm{nDCG@10}$, $\mathrm{MAP}$, and $\mathrm{MRR}$. For the TREC DL series (DL19/DL20/DL21), we follow the default evaluation protocol of Pyserini \cite{lin2021pyserini}: $\mathrm{nDCG@10}$ is computed using graded relevance; while $\mathrm{MAP}$ and $\mathrm{MRR}$ adopt binary relevance, with the relevance threshold set to $l{=}2$ (i.e., qrels with relevance $\geq 2$ are regarded as relevant). This setting avoids counting "marginally relevant" documents (relevance = 1) as relevant under the binary setting, thereby remaining consistent with the commonly used official binary evaluation protocol for the DL benchmarks.
We use the official prebuilt MIRACL-zh index and enable the Chinese analyzer during retrieval.

\subsection{Experimental Results and Analysis}

To evaluate the effectiveness of different query expansion strategies, we conduct comparisons on TREC DL19, DL20, DL21, and MIRACL-zh. The main results are reported in Tables~\ref{tab:dl19Main}--\ref{tab:miraclzhMain}. Throughout this paper, we report gains in terms of absolute improvement (pp).

To verify the statistical significance of performance differences among models, we conduct paired t-tests between the student model after two-stage SFT+DPO training and each baseline method in all comparison experiments. In the tables, $^{\dag}$ indicates that the difference is statistically significant ($p \le 0.05$).

\begin{table}[htbp]
\caption{Comparison of retrieval effectiveness for different query expansion strategies on TREC DL19. $\dag$ indicates a statistically significant difference from SFT+DPO under a paired t-test ($p<0.05$).}
\label{tab:dl19Main}
\centering
\begin{tabular}{lccc}
\toprule
Method & nDCG@10 & MAP & MRR \\
\midrule
\multicolumn{4}{l}{\textbf{BM25 Retriever (DeepSeek Distillation)}} \\
BM25 & 0.5058$^{\dag}$ & 0.3013$^{\dag}$ & 0.7036$^{\dag}$ \\
BM25+Rocchio & 0.5275$^{\dag}$ & 0.3474$^{\dag}$ & 0.6676$^{\dag}$ \\
Original model expansion & 0.6164$^{\dag}$ & 0.4150 & 0.7592$^{\dag}$ \\
Query2Doc & 0.6236 & 0.4177$^{\dag}$ & 0.7971 \\
SFT-only & 0.6405 & 0.4269 & 0.8132 \\
DPO-only & 0.6292 & 0.4334 & 0.8204 \\
SFT+DPO & \textbf{0.6536} & \textbf{0.4421} & \textbf{0.8211} \\
\bottomrule
\end{tabular}
\end{table}

On DL19 (Table~\ref{tab:dl19Main}), the traditional PRF baseline Rocchio achieves a certain improvement over BM25 in nDCG@10 (0.5058$\rightarrow$0.5275), but its MRR slightly decreases (0.7036$\rightarrow$0.6676), indicating that term re-estimation based on the feedback set may improve overall ranking quality while introducing fluctuations in top-ranked hits. In contrast, generative query expansion yields more stable gains. Directly using the original student backbone for expansion already improves nDCG@10 to 0.6164 and reaches 0.4150 on MAP. Query2Doc-style few-shot expansion further improves nDCG@10 to 0.6236 and significantly improves MRR (0.7971). In the distillation and alignment stage, SFT-only outperforms the above direct expansion baselines on all three metrics (0.6405/0.4269/0.8132 in nDCG@10/MAP/MRR), indicating that supervised distillation can effectively transfer the teacher's expansion patterns and improve expansion retrievability. DPO-only achieves performance close to SFT-only on MAP and MRR (0.4334/0.8204), but is slightly lower on nDCG@10 (0.6292), suggesting that preference alignment can strengthen top-rank relevance signals, but its gains are limited without supervised distillation. Finally, the two-stage DPO version achieves the best result on DL19 (nDCG@10 = 0.6536), with absolute improvements of 14.78\,pp, 14.08\,pp, and 11.75\,pp over BM25. At the same time, the two-stage DPO model shows significant differences from most baseline methods on all three metrics (see the $^{\dag}$ marks in the table).

\begin{table}[htbp]
\caption{Comparison of retrieval effectiveness for different query expansion strategies on TREC DL20. $\dag$ indicates a statistically significant difference from SFT+DPO under a paired t-test ($p<0.05$).}
\label{tab:dl20Main}
\centering
\begin{tabular}{lccc}
\toprule
Method & nDCG@10 & MAP & MRR \\
\midrule
\multicolumn{4}{l}{\textbf{BM25 Retriever (DeepSeek Distillation)}} \\
BM25 & 0.4796$^{\dag}$ & 0.2856$^{\dag}$ & 0.6585$^{\dag}$ \\
BM25+Rocchio & 0.4910$^{\dag}$ & 0.3115$^{\dag}$ & 0.6513$^{\dag}$ \\
Original model expansion & 0.5476$^{\dag}$ & 0.3379$^{\dag}$ & 0.6766 \\
Query2Doc & 0.5801 & 0.3633$^{\dag}$ & 0.7120 \\
SFT-only & 0.5797 & 0.3664 & \textbf{0.7258} \\
DPO-only & 0.5846 & 0.3745 & 0.7208 \\
SFT+DPO & \textbf{0.5921} & \textbf{0.3831} & 0.7203 \\
\bottomrule
\end{tabular}
\end{table}

On DL20 (Table~\ref{tab:dl20Main}), Rocchio brings only limited improvement over BM25 on nDCG@10 (0.4796$\rightarrow$0.4910), while generative query expansion yields more substantial and more stable gains. The original model expansion improves nDCG@10 to 0.5476 (an absolute gain of 6.80\,pp over BM25), and Query2Doc (few-shot) further reaches 0.5801, while also improving MAP and MRR. In terms of training strategies, SFT-only is close to Query2Doc on nDCG@10 and MAP (0.5797/0.3664), but achieves the highest MRR in this table (0.7258), indicating that supervised distillation helps improve top-rank hits and ranking stability. DPO-only is slightly better than SFT-only on nDCG@10 and MAP (0.5846/0.3745), showing that retrieval-feedback-based alignment can strengthen the retrieval orientation of expansions; however, its overall gains are still affected by initialization and supervised signals. Finally, the two-stage DPO model achieves the best nDCG@10 and MAP on DL20 (0.5921 and 0.3831, respectively), corresponding to absolute improvements of 11.25\,pp and 9.75\,pp over BM25. At the same time, its MRR also improves by 6.18\,pp over BM25 (0.6585$\rightarrow$0.7203). Overall, these results show that the combination of distillation + preference alignment provides stronger comprehensive gains in both ranking quality and retrieval stability.

\begin{table}[htbp]
\caption{Comparison of retrieval effectiveness for different query expansion strategies on TREC DL21. $\dag$ indicates a statistically significant difference from SFT+DPO under a paired t-test ($p<0.05$).}
\label{tab:dl21Main}
\centering
\begin{tabular}{lccc}
\toprule
Method & nDCG@10 & MAP & MRR \\
\midrule
\multicolumn{4}{l}{\textbf{BM25 Retriever (DeepSeek Distillation)}} \\
BM25 (no expansion) & 0.4458$^{\dag}$ & 0.1708$^{\dag}$ & 0.5060$^{\dag}$ \\
BM25+Rocchio & 0.4544$^{\dag}$ & 0.2151$^{\dag}$ & 0.5442$^{\dag}$ \\
Original model expansion & 0.5186$^{\dag}$ & 0.2543$^{\dag}$ & 0.6659 \\
Query2Doc & 0.5001$^{\dag}$ & 0.2407$^{\dag}$ & 0.6625 \\
SFT-only & 0.5271$^{\dag}$ & 0.2622$^{\dag}$ & 0.6589 \\
DPO-only & 0.5560 & 0.2645$^{\dag}$ & 0.6768 \\
SFT+DPO & \textbf{0.5633} & \textbf{0.2789} & \textbf{0.6849} \\
\bottomrule
\end{tabular}
\end{table}

On DL21 (Table~\ref{tab:dl21Main}), the BM25 baseline is nDCG@10 = 0.4458. Rocchio brings only limited gains on this dataset (0.4458$\rightarrow$0.4544), whereas generative query expansion significantly improves ranking quality. The original model expansion increases nDCG@10 to 0.5186 and significantly improves MAP and MRR. Query2Doc (few-shot) is slightly lower than the original model expansion on nDCG@10 (0.5001), but maintains a relatively high MRR (0.6625), indicating that few-shot prompting is more beneficial for improving top-rank hits, though its overall ranking gains are more sensitive to prompt design and generation style. In terms of training strategies, the SFT student expansion further outperforms the direct expansion baselines on nDCG@10 and MAP (0.5271/0.2622), showing that supervised distillation can stably transfer the retrieval-friendly characteristics of teacher expansions. Notably, DPO-only already reaches 0.5560 on nDCG@10, indicating that preference alignment driven solely by retrieval feedback can also significantly strengthen ranking quality. On top of this, the two-stage DPO student expansion achieves the best results in this table (nDCG@10 = 0.5633, MAP = 0.2789, MRR = 0.6849). Compared with BM25, the absolute improvements of the two-stage DPO model are 11.75\,pp on nDCG@10, 10.81\,pp on MAP, and 17.89\,pp on MRR. Overall, these results indicate that distillation + preference alignment can simultaneously improve overall ranking quality and top-rank relevance on DL21. The differences between most compared methods and the two-stage SFT+DPO model are statistically significant (see the $^{\dag}$ marks in the table).

\begin{table}[htbp]
\caption{Comparison of retrieval effectiveness for different query expansion strategies on MIRACL-zh. $\dag$ indicates a statistically significant difference from SFT+DPO under a paired t-test ($p<0.05$).}
\label{tab:miraclzhMain}
\centering
\begin{tabular}{lccc}
\toprule
Method & nDCG@10 & MAP & MRR \\
\midrule
\multicolumn{4}{l}{\textbf{BM25 Retriever (DeepSeek Distillation)}} \\
BM25 & 0.1801$^{\dag}$ & 0.1486$^{\dag}$ & 0.2207$^{\dag}$ \\
BM25+Rocchio & 0.1740$^{\dag}$ & 0.1451$^{\dag}$ & 0.2058$^{\dag}$ \\
Original model expansion & 0.3302 & 0.2720$^{\dag}$ & 0.3853$^{\dag}$ \\
Query2Doc & 0.3232$^{\dag}$ & 0.2689$^{\dag}$ & 0.3877$^{\dag}$ \\
SFT-only & 0.3251 & 0.2728 & 0.3920$^{\dag}$ \\
DPO-only & 0.3262 & 0.2691$^{\dag}$ & 0.3820$^{\dag}$ \\
SFT+DPO & \textbf{0.3366} & \textbf{0.2807} & \textbf{0.4096} \\
\bottomrule
\end{tabular}
\end{table}

On MIRACL-zh (Table~\ref{tab:miraclzhMain}), Rocchio does not improve over BM25; instead, it slightly degrades all three metrics (nDCG@10: 0.1801$\rightarrow$0.1740, MAP: 0.1486$\rightarrow$0.1451, MRR: 0.2207$\rightarrow$0.2058). This suggests that traditional PRF based on term reweighting is less effective on this Chinese benchmark. In contrast, generative query expansion brings substantial gains. The original model expansion improves nDCG@10, MAP, and MRR to 0.3302, 0.2720, and 0.3853, respectively, showing that paragraph-level expansion can greatly strengthen query representation in the Chinese retrieval setting. Query2Doc achieves comparable performance (0.3232/0.2689/0.3877), indicating that few-shot prompting remains a competitive direct-expansion strategy on MIRACL-zh, even when the few-shot demonstrations are reused from the English setting. 
In terms of training strategies, SFT-only further improves MAP and MRR over the direct expansion baselines (0.2728/0.3920), suggesting that supervised distillation helps the student absorb retrieval-friendly generation patterns from the teacher. DPO-only performs similarly to the direct expansion baselines, but remains weaker than SFT-based training overall, indicating that preference alignment alone is insufficient without supervised initialization. Finally, the two-stage SFT+DPO student expansion achieves the best results on all three metrics, reaching 0.3366 nDCG@10, 0.2807 MAP, and 0.4096 MRR. Compared with BM25, this corresponds to absolute gains of +15.65\,pp on nDCG@10, +13.21\,pp on MAP, and +18.89\,pp on MRR.  
Overall, the MIRACL-zh results show that the Chinese benchmark does not favor Rocchio-style lexical PRF; instead, distillation combined with preference alignment provides the most consistent gains in both ranking quality and top-rank relevance.
The $^{\dag}$ marks indicate statistically significant differences from the best result (SFT+DPO).

Across the four benchmarks, we observe the following. (1) Compared with sparse expansion based on term reweighting (Rocchio), generative query expansion more consistently improves nDCG@10 and MRR across datasets, and on MIRACL-zh the advantage is particularly clear, indicating that paragraph-level semantic generation is more effective than lexical PRF for strengthening query representation in the Chinese retrieval setting. (2) SFT distillation effectively inherits the teacher's expansion patterns without changing model size, and generally outperforms direct expansion baselines that rely only on prompting, demonstrating the contribution of supervised distillation to expansion stability and retrievability. (3) Further introducing DPO alignment on top of SFT consistently brings additional nDCG@10 gains across all four benchmarks, namely +1.31\,pp, +1.24\,pp, +3.62\,pp, and +1.15\,pp on DL19, DL20, DL21, and MIRACL-zh, respectively, while also improving or maintaining competitive MAP/MRR on most datasets. Overall, these results suggest that retrieval-metric-driven preference alignment is particularly helpful for improving ranking quality after supervised distillation, and that this benefit extends from English TREC DL benchmarks to the Chinese MIRACL-zh benchmark as well.

\subsection{Efficiency and Resource Cost Analysis}
\label{sec:efficiency}

\subsubsection{End-to-End Generation Latency}

Beyond effectiveness metrics, a key bottleneck for deploying generative query expansion in practice is inference latency. To quantitatively evaluate the generation cost of models with different scales and methods, we conduct comparative experiments. The latency measurement includes two categories: (1) \textbf{local inference}: end-to-end generation time of open-source models is measured on a single RTX PRO 6000 (96GB) GPU; the software environment is PyTorch 2.9.1 / Python 3.12 (Ubuntu 22.04) / CUDA 12.8; the CPU is a 22 vCPU Intel Xeon Platinum 8470Q with 110GB memory. The same decoding configuration is used during generation, and timing is measured with batch size = 2. (2) \textbf{API end-to-end response time}: for DeepSeek-V3.1 (685B), queries are sent one by one through the official API, and the total elapsed time from request initiation to receiving the full expansion is recorded.

\begin{table}[htbp]
\caption{Efficiency comparison of query expansion generation on DL19 and DL20 under different models and prompting settings.}
\label{tab:efficiency_gen}
\centering
\begin{tabular}{lrr}
\toprule
Setting & Time (s) & s/query \\
\midrule
\multicolumn{3}{l}{\textit{DL19} (Queries = 43)} \\
DeepSeek-V3.1 (685B) API & 203.24 & 4.726 \\
OLMo3.1-32B (Teacher, local) & 98.77 & 2.297 \\
Query2Doc (Qwen3-4B, local few-shot) & 17.86 & 0.415 \\
SFT+DPO (Qwen3-4B, local) & \textbf{4.44} & \textbf{0.103} \\
\midrule
\multicolumn{3}{l}{\textit{DL20} (Queries = 54)} \\
DeepSeek-V3.1 (685B) API & 262.70 & 4.865 \\
OLMo3.1-32B (Teacher, local) & 122.77 & 2.274 \\
Query2Doc (Qwen3-4B, local few-shot) & 22.17 & 0.410 \\
SFT+DPO (Qwen3-4B, local) & \textbf{5.38} & \textbf{0.100} \\
\bottomrule
\end{tabular}
\end{table}

From Table~\ref{tab:efficiency_gen}, we can draw the following conclusions. First, under the same local hardware, the distilled 4B DPO student model is significantly faster than the 32B teacher model in generation latency. On DL19, the DPO student requires 0.103 s/query, while OLMo-32B requires 2.297 s/query, corresponding to a speedup of approximately $2.297/0.103 \approx 22.25\times$. On DL20, the speedup is similarly about $2.274/0.100 \approx 22.83\times$. These results show that after transferring the query expansion capability to a smaller model via knowledge distillation and preference alignment, the end-to-end latency of online expansion generation can be significantly reduced, making the method better suited to the low-latency requirements of practical retrieval systems. At the same time, within our framework, the teacher model is mainly used to generate training data offline; during online serving, only the student model needs to be deployed, and thus the overall online inference cost is determined by the student model.

Second, Table~\ref{tab:efficiency_gen} also reports the API end-to-end response time of DeepSeek-V3.1 (685B) as an engineering reference. Its zero-shot latency is 4.726 s/query and 4.865 s/query on DL19 and DL20, respectively. Compared with the locally deployed 4B student model, the end-to-end latency gap is about $4.726/0.103 \approx 45.9\times$ on DL19 and $4.865/0.100 \approx 48.6\times$ on DL20. 
We report API latency as a realistic end-to-end reference for online invocation, to support the claim that online deployment of large models is usually difficult to reconcile with low-latency requirements.

Finally, regarding the cost difference caused by prompting strategies, Query2Doc-style few-shot prompting on the local 4B model significantly increases the input length due to the inclusion of multiple in-context examples, thereby noticeably increasing generation time: 0.415 s/query on DL19 and 0.410 s/query on DL20. Compared with the student model after DPO, it is about $0.415/0.103 \approx 4.02\times$ slower on DL19 and $0.410/0.100 \approx 4.10\times$ slower on DL20. This indicates that although Query2Doc has the advantage of being training-free, its online generation cost is higher.

\subsubsection{Peak GPU Memory Usage}

\begin{figure}[htbp]
\centering
\includegraphics[width=0.7\linewidth]{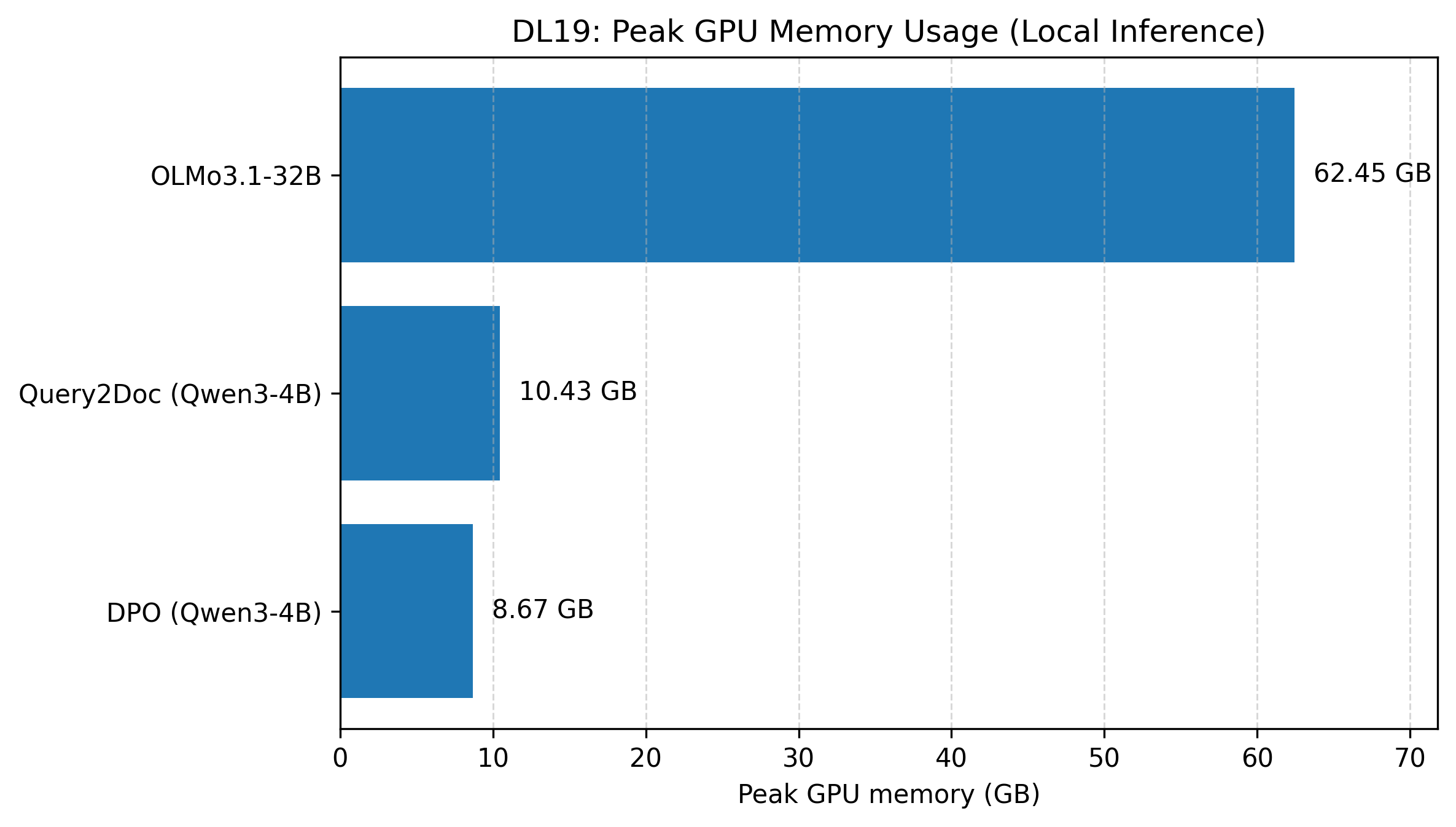}
\caption{Peak GPU memory usage for query expansion generation under different settings on DL19 (local inference).}
\label{fig:dl19_gpu_mem}
\end{figure}

In addition to inference latency, GPU memory usage is another key constraint for the online deployment of generative query expansion. We measure the peak GPU memory cost of local inference on DL19 and compare the memory requirements of the 32B teacher model (OLMo3.1-32B), 4B student models, and the few-shot baseline. The results are shown in Fig.~\ref{fig:dl19_gpu_mem}. As can be seen, the peak GPU memory usage of the 32B teacher model is about 63.9GB, which is substantially higher than the 8.9--10.7GB range of the 4B models. Among them, the student model after our two-stage DPO training achieves the lowest generation latency while maintaining the smallest memory footprint (about 8.9GB), indicating that the distilled small model is more suitable for resource-constrained online deployment scenarios. It should be noted that the backend inference hardware and parallelization strategy of the DeepSeek API are not externally visible, so its GPU memory usage cannot be reported under the same measurement protocol.

\subsection{Extended Experiments and Ablation Analysis}

\subsubsection{Analysis of DeepSeek Teacher Expansion Quality and Prompting Strategies}

Table~\ref{tab:deepseekPrompt} reports the impact of expansion texts generated by the teacher model DeepSeek (685B) under two prompting configurations (zero-shot / few-shot) on BM25 retrieval effectiveness. Overall, the expansions generated by the teacher model bring stable ranking gains on all four datasets, indicating that a high-capacity LLM has the potential to directly generate retrieval-friendly expansion text. The relative advantage of different prompting strategies depends on the dataset distribution: on DL19 and DL20, few-shot improves nDCG@10 over zero-shot by 3.10\,pp and 1.70\,pp, respectively; on DL21, however, zero-shot is slightly better, suggesting that the example style used in few-shot prompting does not yield consistent gains on all tasks. On MIRACL-zh, few-shot also performs slightly better than zero-shot, improving nDCG@10 from 0.3777 to 0.3837 (+0.60\,pp). Notably, this gain is obtained even though the few-shot exemplars are kept in their original English form while the target queries are in Chinese, suggesting that the teacher's expansion behavior is reasonably robust to prompt-language mismatch. At the same time, the teacher still achieves stronger results than the distilled student on MIRACL-zh, indicating that the Chinese setting remains a meaningful and non-trivial target for distillation.

\begin{table}[htbp]
\caption{Teacher expansion quality under different prompting configurations (DeepSeek).}
\label{tab:deepseekPrompt}
\centering
\begin{tabular}{lcccc}
\toprule
Dataset & Prompting & nDCG@10 & MAP & MRR \\
\midrule
\multirow{2}{*}{DL19}
& zero-shot & 0.6741 & 0.4588 & 0.8184 \\
& few-shot  & 0.7051 & 0.4912 & 0.8510 \\
\midrule
\multirow{2}{*}{DL20}
& zero-shot & 0.6417 & 0.4316 & 0.7931 \\
& few-shot  & 0.6587 & 0.4570 & 0.8568 \\
\midrule
\multirow{2}{*}{DL21}
& zero-shot & 0.5815 & 0.2953 & 0.6865 \\
& few-shot  & 0.5384 & 0.2865 & 0.6798 \\
\midrule
\multirow{2}{*}{MIRACL-zh}
& zero-shot & 0.3777 & 0.3232 & 0.4470 \\
& few-shot  & 0.3837 & 0.3275 & 0.4513 \\
\bottomrule
\end{tabular}
\end{table}

Furthermore, the best results of the teacher model on multiple datasets (e.g., nDCG@10 = 0.7051 on DL19) are overall higher than the expansion effectiveness of the small model under the setting of prompt-only without training, which reflects the advantage of large models in term coverage, semantic elaboration, and paragraph organization. Therefore, using teacher expansions as supervision signals to distill the student model is well motivated: the goal is to preserve as much of the teacher's retrieval gains as possible without relying on online calls to a large model, while simultaneously reducing inference cost and deployment barriers.

Combined with the results in Tables~\ref{tab:dl19Main}--\ref{tab:miraclzhMain}, we can observe that after SFT+DPO distillation, the 4B student model is already able to approach the teacher model's retrieval effectiveness. For example, on DL19, the student model reaches nDCG@10 = 0.6536, which is about 96.96\% of the teacher's zero-shot result (0.6741).

\subsubsection{Robustness Across Different Teacher Models}
\label{sec:otherOlmo32Teacher}

In addition to experiments with the DeepSeek-685B teacher model, we further examine the transferability and robustness of the proposed distillation framework under different teacher models. Specifically, we use the open-source large model OLMo-3.1-32B-Instruct\footnote{\url{https://huggingface.co/allenai/Olmo-3.1-32B-Instruct}} as the teacher model, following the same data construction pipeline as in the main experiments. For each query, paragraph-level expansions are generated under both zero-shot and few-shot prompts, and the same student backbone Qwen3-4B-Instruct-2507 is trained accordingly. The student model is then evaluated under three settings: (1) SFT only, (2) DPO-only, and (3) the full two-stage setting of SFT followed by DPO. The construction of the DPO training data is the same as in the main experiments, except that the data are generated by OLMo-3.1-32B. Evaluation still uses the BM25 retrieval setting on DL19 and the same retrieval metrics (nDCG@10/MAP/MRR), ensuring comparability with the main experiments.

\begin{table}[htbp]
\caption{Retrieval performance (BM25) on TREC DL19 with OLMo-32B as the teacher model.}
\label{tab:DL19_OlmoTeacher}
\centering
\begin{tabular}{lccc}
\toprule
Method & nDCG@10 & MAP & MRR \\
\midrule
OLMo-32B expansion (zero-shot) & 0.6444 & 0.4180 & 0.8118 \\
OLMo-32B expansion (few-shot)  & 0.6521 & \textbf{0.4401} & \textbf{0.8595} \\
\midrule
SFT-only & 0.6395 & 0.4305 & 0.8482 \\
DPO-only & 0.6277 & 0.4326 & 0.8587 \\
SFT+DPO  & \textbf{0.6599} & 0.4388 & 0.8318 \\
\bottomrule
\end{tabular}
\end{table}

Table~\ref{tab:DL19_OlmoTeacher} presents the retrieval results on TREC DL19 (BM25) for different expansion and training strategies when OLMo-32B is used as the teacher model. First, directly generated expansions from OLMo-32B already bring clear gains, and few-shot outperforms zero-shot on MAP and MRR (0.4401/0.8595 vs. 0.4180/0.8118), indicating that in-context examples help generate more retrieval-oriented expansion expressions. Second, after distilling from OLMo teacher data, the student model still maintains effectiveness close to teacher expansions overall: both the SFT student and DPO-only are competitive on MAP/MRR, which indicates that the proposed training pipeline is not sensitive to the source of the teacher and has cross-teacher transferability.

Finally, the two-stage DPO student expansion achieves the best nDCG@10 in this table (0.6599), even slightly higher than the teacher expansion (few-shot: 0.6521). This phenomenon does not mean that the student model fully surpasses the teacher in generation capability; rather, it is more likely due to the difference in alignment objectives. Teacher expansions are mainly influenced by prompting and language-generation preferences, and are not necessarily optimal for BM25 ranking metrics; in contrast, DPO directly constructs preference supervision from retrieval metrics, encouraging the student to generate expansions with fuller term coverage and stronger matching signals, thereby obtaining higher ranking gains on this dataset. Overall, this experiment further verifies that the pipeline of teacher generation $\rightarrow$ SFT distillation $\rightarrow$ DPO alignment can be reproduced on open-source teacher models of different scales and architectures, improving the generality and reusability of the method.

\subsubsection{Generalization Across Different Retrievers: Dense Retrieval}

The main experiments in this paper use BM25 as the base retriever, focusing on whether the distillation and alignment of a generative query expansion model can substantially improve effectiveness and reduce inference cost under a traditional sparse retrieval setting. Considering that recent information retrieval systems also widely adopt dense retrievers, we further conduct supplementary experiments on bge-base-en-v1.5\footnote{\url{https://huggingface.co/BAAI/bge-base-en-v1.5}} to verify the generality of the proposed method. Specifically, we use the pre-built BGE Faiss flat index for MS MARCO v1 passage provided by Pyserini, and adopt its recommended query prefix ("Represent this sentence for searching relevant passages:") together with $\ell_2$ normalization for retrieval.

\begin{table}[htbp]
\caption{Dense retrieval results with BGE-base-en-v1.5 on TREC DL19 and DL20. $\dag$ indicates a statistically significant difference from SFT+DPO under a paired t-test ($p<0.05$).}
\label{tab:bge_dl19_dl20}
\centering
\begin{tabular}{lccc}
\toprule
Method & nDCG@10 & MAP & MRR \\
\midrule
\multicolumn{4}{l}{\textbf{DL19 (BGE-base-en-v1.5)}} \\
BGE (raw) & 0.7016 & 0.4485$^{\dag}$ & 0.9419 \\
Original model expansion & 0.6680$^{\dag}$ & 0.4521$^{\dag}$ & 0.9034 \\
Query2Doc & 0.7111 & 0.4779 & 0.9535 \\
SFT-only & 0.7004 & 0.4725 & \textbf{0.9536} \\
DPO-only & 0.6886 & 0.4658 & 0.9420 \\
SFT+DPO & \textbf{0.7226} & \textbf{0.4886} & 0.9535 \\
\midrule
\multicolumn{4}{l}{\textbf{DL20 (BGE-base-en-v1.5)}} \\
BGE (raw) & 0.6768$^{\dag}$ & 0.4628$^{\dag}$ & 0.9256$^{\dag}$ \\
Original model expansion & 0.6571$^{\dag}$ & 0.4475$^{\dag}$ & 0.9189 \\
Query2Doc & 0.6697$^{\dag}$ & 0.4537$^{\dag}$ & 0.9091 \\
SFT-only & 0.6976 & 0.4835 & 0.9321 \\
DPO-only & 0.6825$^{\dag}$ & 0.4758$^{\dag}$ & 0.9204 \\
SFT+DPO & \textbf{0.7057} & \textbf{0.4944} & \textbf{0.9396} \\
\bottomrule
\end{tabular}
\end{table}

Table~\ref{tab:bge_dl19_dl20} reports dense retrieval results with BGE-base-en-v1.5 on TREC DL19 and DL20.
Based on the Dense (BGE) part, we obtain the following observations.

\textbf{Under a strong dense baseline, direct expansion does not necessarily help.}  
Taking DL19 as an example, BGE(raw) already achieves strong results (nDCG@10 = 0.7016, MAP = 0.4485, MRR = 0.9419), but "original model expansion" instead causes a clear drop in nDCG@10 (0.6680). This indicates that under dense retrieval, expansion content that introduces drift or redundancy may weaken representation quality and matching stability.

\textbf{Two-stage distillation still provides stable gains under dense retrievers.}  
More importantly, the proposed method continues to bring consistent improvements in the BGE setting. On DL19, our SFT+DPO version reaches \textbf{0.7226} nDCG@10 and \textbf{0.4886} MAP; on DL20, it further reaches \textbf{0.7057} nDCG@10 and \textbf{0.4944} MAP. Compared with BGE(raw), our method improves by +0.0210 (nDCG@10) and +0.0401 (MAP) on DL19, and by +0.0289 and +0.0316 on DL20, respectively. This shows that the training objective of constructing preferences from retrieval feedback and aligning generation behavior is also effective for dense retrieval.

\textbf{The instability of DPO-only.}  
From the results, DPO-only exhibits a certain degree of instability across datasets and retrievers. For example, on DL19 under the dense retriever, DPO-only obtains 0.6886/0.4658 on nDCG@10/MAP, which is clearly lower than the two-stage SFT+DPO model. On DL20, the two-stage SFT+DPO model also outperforms DPO-only on all three metrics, improving nDCG@10 from 0.6825 to 0.7057, MAP from 0.4758 to 0.4944, and MRR from 0.9204 to 0.9396. These results suggest that, without the prior constraint provided by SFT, DPO is more susceptible to noisy or weak preference signals, leading to performance fluctuations across different datasets and retrievers. In contrast, first distilling the teacher's retrieval-friendly expansion distribution via SFT and then applying DPO for retrieval-oriented alignment is overall more robust.

In summary, the supplementary experiments show that although the main experiments of this paper adopt BM25 to match traditional and reproducible retrieval settings, the proposed distillation and alignment framework does not depend on a specific retriever type, and also yields stable gains with the strong dense baseline BGE-base-en-v1.5.

\subsubsection{Effect of Different LoRA Ranks}

\begin{figure}[htbp]
\centering
\includegraphics[width=0.65\linewidth]{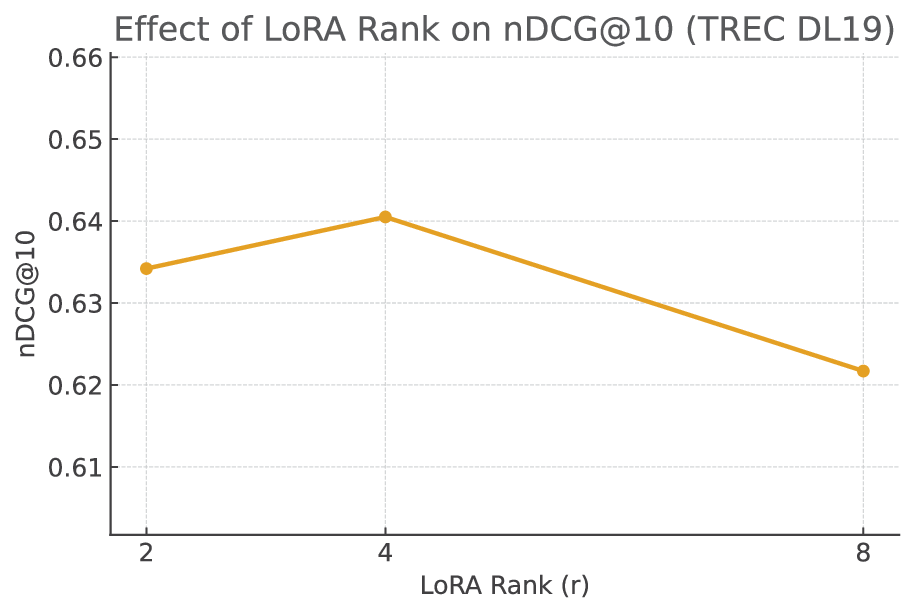}
\caption{Effect of LoRA rank on nDCG@10 performance of the model after SFT on TREC DL19.}
\label{fig:lora_rank_ndcg}
\end{figure}

We report the effectiveness on TREC DL19 when the student model is first fine-tuned from teacher data under different LoRA ranks, as shown in Fig.~\ref{fig:lora_rank_ndcg}. Under different LoRA ranks, the retrieval performance of the student model exhibits a trend in which a medium rank is optimal. When the rank is 2, the model can already reproduce the major semantic structure of teacher expansions. Increasing the rank to 4 further improves nDCG@10. However, when the rank is further increased to 8, nDCG@10 drops noticeably. This suggests that an excessively large rank may introduce redundant parameters, leading to unstable training and slight overfitting.

\subsubsection{Sensitivity to the Preference Threshold $\Delta$}

\begin{figure}[htbp]
\centering
\includegraphics[width=0.65\linewidth]{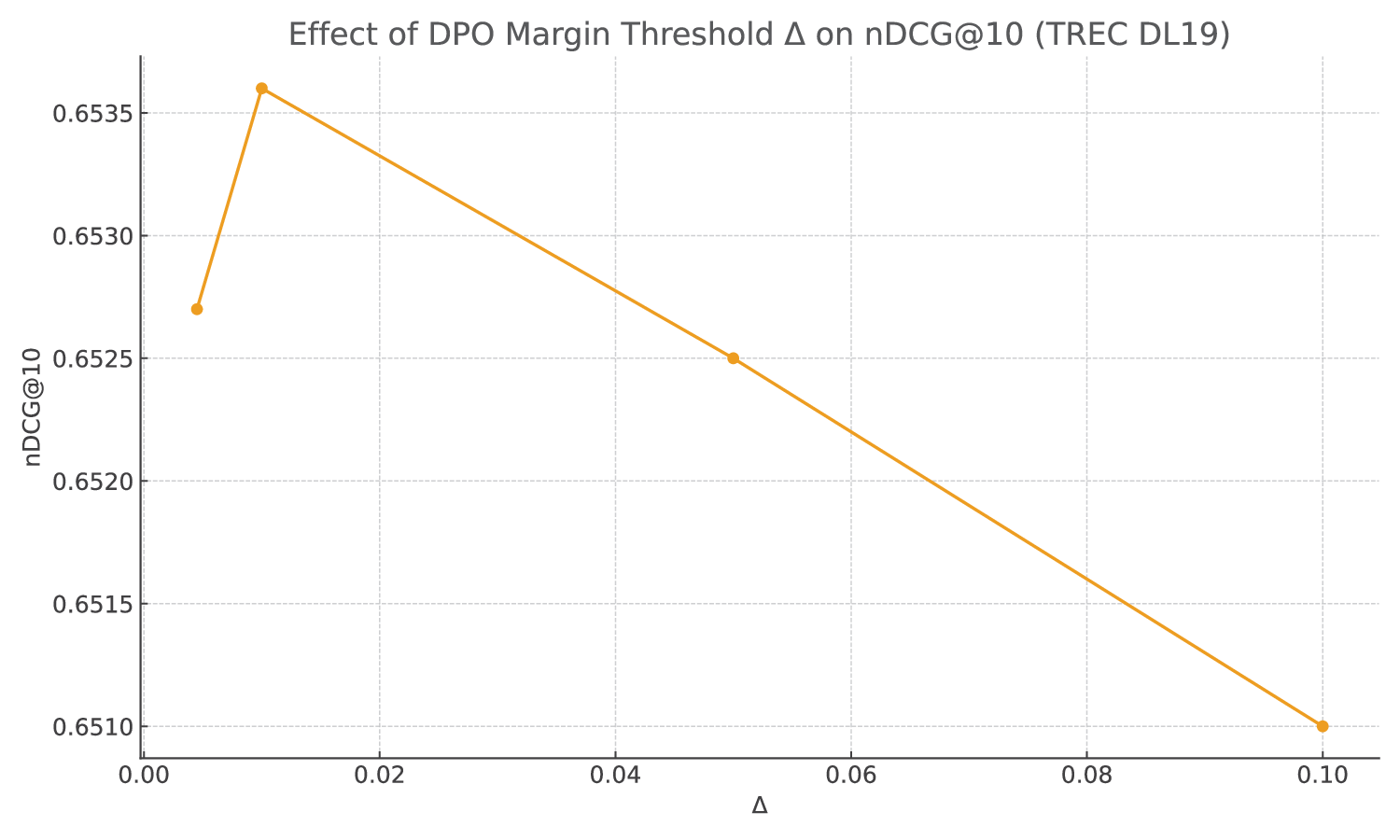}
\caption{Effect of the DPO preference-margin threshold $\Delta$ on nDCG@10 performance on TREC DL19.}
\label{fig:delta_ndcg}
\end{figure}

In this work, the preference pairs used in DPO come from the relative quality difference between two generated expansions, where the preference threshold $\Delta$ is used to retain only those preference pairs with sufficiently clear advantages, so as to reduce training interference caused by noise. We test different values of $\Delta$ (0.005, 0.01, 0.05, 0.1) after DPO preference alignment, and the results are shown in Fig.~\ref{fig:delta_ndcg}. When $\Delta$ is small (e.g., 0.005 and 0.01), the model achieves relatively high nDCG@10, indicating that a small threshold can retain a sufficient number of preference samples and thus provide rich optimization signals for DPO. Among them, $\Delta = 0.01$ performs slightly better, achieving the highest nDCG@10 and exhibiting more stable training. In contrast, $\Delta = 0.005$ is slightly lower, possibly because the differences between preference pairs are too small, making the reward signal less clear and thus weakening effective guidance for generation preferences.

As the threshold increases to 0.05 and 0.1, the number of available preference samples decreases substantially, the training signal becomes sparse, and model performance correspondingly declines. However, it should be noted that nDCG@10 remains above 0.65 for all four thresholds, with only limited variation, indicating that the DPO preference alignment process itself has a certain degree of stability and robustness. Considering these results together, we adopt $\Delta = 0.01$ in the main experiments.

\subsubsection{Ablation on SFT Distillation Data Composition: E1+E2, E1-only, and E2-only}
\label{sec:ablation-e1e2}

In distillation-based query expansion, the teacher model can generate multiple expansions for the same query under different prompting strategies. Among them, zero-shot expansion (E1) is usually more concise and more oriented toward general semantic coverage and key concept completion; few-shot expansion (E2), by contrast, is more structured and closer to the example style, often containing richer details and terminology. Since the two types of expansions may be complementary in information density, expression style, and concept coverage, we further investigate the effect of SFT-stage distillation data composition on student model training.

\begin{table}[htbp]
\caption{Ablation on SFT distillation data composition (DL19 + BM25).}
\label{tab:ablation_e1e2_dl19_bm25}
\centering
\begin{tabular}{lccc}
\toprule
Method (SFT Data) & nDCG@10 & MAP & MRR \\
\midrule
SFT(E1+E2)   & \textbf{0.6405} & \textbf{0.4269} & \textbf{0.8132} \\
SFT(E1-only) & 0.6275 & 0.4222 & 0.7758 \\
SFT(E2-only) & 0.6230 & 0.4089 & 0.7585 \\
\bottomrule
\end{tabular}
\end{table}

To this end, we design a data-composition ablation and present the results in Table~\ref{tab:ablation_e1e2_dl19_bm25}, comparing three SFT training settings:  
(1) \textbf{E1+E2}: for each query, both teacher expansions E1 and E2 are retained and jointly used for SFT;  
(2) \textbf{E1-only}: only zero-shot expansions are used as SFT training data;  
(3) \textbf{E2-only}: only few-shot expansions are used as SFT training data.  
Except for the training data composition, all other settings remain unchanged, including the student backbone, LoRA parameter settings, training hyperparameters, and decoding strategy (as well as the subsequent DPO alignment configuration). This ablation is intended to verify whether expansions from different prompt sources provide complementary gains during SFT, and how performance changes when relying on only a single source.

From Table~\ref{tab:ablation_e1e2_dl19_bm25}, we can observe that \textbf{SFT trained on the combined E1+E2 data achieves the best results on all three metrics}, indicating that zero-shot and few-shot expansions are complementary during the distillation stage. E1 provides broader concept coverage and more stable general expressions, whereas E2 contributes richer details and stronger structural expression, and together they improve the expansion quality of the student model. In contrast, using only a single source of data (E1-only or E2-only) leads to consistent performance drops. Among them, the decline of E2-only is more pronounced, suggesting that few-shot-style expansions are not sufficient to fully replace the contribution of zero-shot expansions in terms of coverage and robustness.

\section{Conclusion}
This paper presents an integrated ``teacher--student--preference-alignment'' framework for query expansion. Specifically, DeepSeek-generated expansions are used as teacher signals, and a compact student model is first distilled via supervised fine-tuning (SFT), and then further aligned with retrieval objectives through Direct Preference Optimization (DPO) using preference pairs constructed from retrieval metric differences (nDCG@10). In this way, the prompt-induced expansion capability of a strong teacher model is transferred into a directly deployable small model, enabling retrieval-friendly paragraph-level expansion without relying on few-shot demonstrations at inference time.

Experiments on TREC DL19/20/21 and MIRACL-zh show that, compared with no expansion, BM25+Rocchio, and direct student-side prompting baselines, the proposed method brings consistent improvements in nDCG@10, MAP, and MRR, while the DPO stage further amplifies the gains introduced by SFT. In particular, on TREC DL19, the distilled Qwen3-4B student reaches about 97\% of the teacher model's nDCG@10 performance, demonstrating strong practical value under a low-cost deployment setting. On MIRACL-zh, the proposed framework also yields substantial gains over BM25 and other student-side expansion baselines, showing that the method remains effective beyond the original English MS MARCO-style setting and can be extended to Chinese retrieval scenarios as well.


\bibliographystyle{ACM-Reference-Format}
\bibliography{sample-base}










\end{document}